\documentclass[twocolumn,preprintnumbers]{revtex4}

\usepackage{graphicx}
\usepackage{epsf}
\usepackage{amsmath,amssymb}
\newcommand{\bvec}{\boldsymbol}

\begin{document}
\title{Proton and neutron correlations in $^{10}$B}
\author{Yoshiko Kanada-En'yo}
\affiliation{Department of Physics, Kyoto University, Kyoto 606-8502, Japan}
\author{Hiroyuki Morita}
\affiliation{Department of Physics, Kyoto University, Kyoto 606-8502, Japan}
\author{Fumiharu Kobayashi}
\affiliation{Department of Physics, Niigata University, Niigata 950-2181, Japan}
\begin{abstract}
We investigate positive-parity states of $^{10}$B 
with the calculation of antisymmetrized molecular dynamics
focusing on $pn$ pair correlations. 
We discuss effects of the spin-orbit interaction
on energy spectra and $pn$ correlations of the $J^\pi T=1^+_10$, $=3^+_10$, and 
$0^+_11$ states.
The $1^+_10$ state has almost no energy gain of the spin-orbit interaction,
whereas the $3^+_10$ state gains the spin-orbit interaction energy largely
to come down to the ground state. 
We interpret a part of the two-body spin-orbit interaction 
in the adopted effective interactions
as a contribution of the genuine $NNN$ force, and find it to be essential 
for the level ordering of 
the $3^+_10$ and $1^+_10$ states in $^{10}$B. 
We also apply a $2\alpha+pn$ model to discuss effects of the spin-orbit interaction on
$T=0$ and $T=1$ $pn$ pairs around the 2$\alpha$ core.
In the spin-aligned $J^\pi T=3^+0$ state, the spin-orbit interaction affects the $(ST)=(10)$ pair attractively
and keeps the pair close to the core,  
whereas, in the $1^+0$ state, it gives a minor effect to the $(ST)=(10)$ pair.  In the $0^+1$ state, 
the $(ST)=(01)$ pair is somewhat dissociated by the spin-orbit 
interaction.
\end{abstract}

\maketitle
\section{Introduction}
In the progress of experimental researches on proton-rich nuclei, 
the interest of proton and neutron $(pn)$ pair correlations has been revived
in these years.
In the study of $pn$ pairing, 
the competition between isoscalar $T=0$ pairing and isovector $T=1$ pairing
is one of the essential problems in $Z\sim N$ nuclei 
\cite{Goodman79,Engel:1996sh,Satula:1996dc,Poves:1998br,Goodman:1998zz,Kaneko:2004ya,Baroni:2009eh,Bertsch:2009xz,Gezerlis:2011rh,Sagawa:2012ta}. 
The nuclear interaction in a free space is
more attractive in the $T=0$ spin-triplet 
even ($^3$E) channel than in the $T=1$ spin-singlet even ($^1$E) channel
as known from the bound state, deuteron, formed by two nucleons in the $^3$E channel. 
However, at the nuclear surface and in nuclear medium, the competition between 
$T=0$ and $T=1$ $pn$ pairs occurs.  
Because of the stronger $^3$E interaction than the $^1$E interaction, 
it is naively expected that the deuteron-like $T=0$ pair is more favored 
than the $T=1$ pair as seen in the ground state spin, $J^\pi T=1^+0$,
of $^{6}$Li and $^{18}$F.
However, 
the $T=1$ pair is often favored rather
than the $T=0$ pair in medium- and heavy-mass regions
as seen in the ground state spins of $Z=N=$odd nuclei
because the spin-orbit mean potential favors the $T=1$ pair
\cite{Macchiavelli:1999kf}.  Moreover, the spin-orbit potential favors 
a spin-aligned $T=0$ $pn$ pair \cite{Cederwall:2011dt,Zerguine:2011zb,Qi:2011yj}.
These facts indicate that the spin-orbit interaction plays an important role
in the competition between $T=0$ and $T=1$ $pn$ pairs in nuclear
systems.

Investigations of $Z=N={\rm odd}$ nuclei are helpful to 
understand features of $pn$ pairs at the nuclear surface.
Based on a three-body picture of a core nucleus with 
two valence nucleons, one can discuss the 
competition between $T=0$ and $T=1$ $pn$ pairs 
from the ordering of $J^\pi T=1^+0$ and  
$0^+1$ states. For example, 
$^6$Li and $^{18}$F have the $J^\pi T=1^+0$ ground states and 
the $J^\pi T=0^+1$ excited states indicating that the $T=0$ pair is favored 
rather than the $T=1$ pair. On the other hand, in $^{42}$Sc, the ground state is 
$J^\pi T=0^+1$ because the $T=1$ pair is favored by 
the spin-orbit potential at the surface of the $^{40}$Ca core 
as discussed by Tanimura {\it et al.} 
based on a three-body model calculation \cite{Tanimura:2013cea}.

In the previous paper  \cite{Kanada-En'yo:2014oaa}, two of the authors, Kanada-En'yo and 
Kobayashi, discussed effects of the spin-orbit interaction on $pn$ pairs at the surface 
of $^{16}$O in $^{18}$F based on an $^{16}$O+$pn$ model, 
and found that the level structure of  $J^\pi T=1^+0$, 
$0^+1$, and $3^+0$ states is affected by the strength of 
the spin-orbit interaction. 
Namely, the spin-orbit interaction reduces 
the $T=1$ pair energy in the $0^+1$ state, and 
it largely contributes to the energy of a spin-aligned $T=0$ $pn$ pair attractively to lower the 
$3^+0$ energy, whereas it gives a minor effect 
to the $T=0$ pair energy in the $1^+0$ state.

In $^{10}$B, the ground state is the $3^+0$ state and the first excited state is the 
$1^+0$ state at $E_x=0.72$ MeV. 
Based on a $2\alpha$+$pn$ picture, this fact indicates that $^{10}$B is an 
interesting system in which 
the level inversion between the $1^+0$ state having a $T=0$ pair in an $S$-wave
(a pair moving in the total-angular-momentum $L=0$ state around the core)
and the $3^+0$ state having a spin-aligned $T=0$ pair 
(a pair moving in a total-angular-momentum $L=2$($D$-wave) state around the core)
occurs.
In these years, {\it ab initio} calculations using the 
no-core shell model (NCSM) approach with realistic nuclear forces based on the chiral  perturbation theory
\cite{Machleidt:2011zz} have been achieved for $A\sim 10$ nuclei \cite{Navratil:2007we}. 
The NCSM calculations with effective interactions derived from the chiral 
two-body ($NN$) and three-body ($NNN$) nuclear forces \cite{Epelbaum:2002vt,Entem:2003ft}
describe well the experimental low-lying spectra of $^{10}$B and
show that the $NNN$ force is 
essential to reproduce the ordering of the $3^+0$ and $1^+0$ states 
in $^{10}$B. Recently, Kohno pointed out that the $NNN$ force 
provides an attractive contribution to the effective two-body spin-orbit interaction 
in nuclear medium
\cite{Kohno:2012vj}.
Therefore, it is expected that the $NNN$ force may also contribute to
$pn$ pairs in nuclei through the effective spin-orbit interaction. 

In this paper, we investigate structure of $^{10}$B and clarify 
effects of the spin-orbit interaction on $T=0$ and $T=1$ $pn$ pairs
based on the calculation of antisymmetrized molecular 
dynamics (AMD) \cite{KanadaEnyo:1995tb,KanadaEnyo:1995ir,AMDsupp}
using phenomenological effective nuclear interactions.
The AMD method is a model for structure studies
and has been proved to be one of the successful methods for light nuclei, 
in particular, to describe cluster structures of ground and excited states.
For instance, $2\alpha$ cluster structures of neutron-rich Be isotopes are described 
systematically with the AMD calculations \cite{KanadaEnyo:1995tb,KanadaEn'yo:2012bj}. 
We calculate $^{10}$B with the AMD method 
and find that $2\alpha$ cluster core is 
formed in $^{10}$B. We discuss the role of the spin-orbit interaction in energy spectra
and features of a $pn$ pair around the $2\alpha$ core in $^{10}$B. Moreover, we associate a part of 
the effective two-body spin-orbit interaction with the $NNN$ force 
based on the Kohno's evaluation, and discuss its effect on the $^{10}$B energy spectra.
We also discuss $^6$Li spectra having a $pn$ pair around an $\alpha$ core for comparison.

This paper is organized as follows. In Sec.~\ref{sec:formulation}, we explain the formulation of the AMD method. In Sec.~\ref{sec:results}, the calculated results for $^{10}$B are shown. 
We discuss effects of the spin-orbit interaction on the  $^{10}$B energy spectra 
based on the AMD result in Sec.~\ref{sec:discussion}.
In Sec.~\ref{sec:2alpha-pn}, we perform an analysis using a $2\alpha+pn$ model to discuss 
effects of the spin-orbit interaction on the $pn$ pair around the $2\alpha$ core.
A summary is given in Sec.~\ref{sec:summary}.

\section{Formulation of AMD and effective nuclear interactions}\label{sec:formulation}

\subsection{AMD method}
We apply the method of the variation after parity and total-angular-momentum projections of the 
AMD model (AMD+VAP) \cite{KanadaEn'yo:1998rf,KanadaEn'yo:2003ue} 
to obtain $A$-nucleon wave functions for the ground and excited states
of a nucleus with the mass number $A$. We here briefly explain the formulation of the present AMD calculation.

An AMD wave function is given by a Slater determinant,
\begin{equation}
 \Phi_{\rm AMD}({\bf Z}) = \frac{1}{\sqrt{A!}} {\cal{A}} \{
  \varphi_1,\varphi_2,...,\varphi_A \},
\end{equation}
where  ${\cal{A}}$ is the antisymmetrizer, and the $i$th single-particle wave function 
is written by a product of
spatial($\phi_i$), intrinsic spin($\chi_i$) and isospin($\tau_i$) 
wave functions as
\begin{eqnarray}
 \varphi_i&=& \phi_{{\bf X}_i}\chi_i\tau_i,\\
 \phi_{{\bf X}_i}({\bf r}_j) & = &  \left(\frac{2\nu}{\pi}\right)^{4/3}
\exp\bigl\{-\nu({\bf r}_j-\frac{{\bf X}_i}{\sqrt{\nu}})^2\bigr\},
\label{eq:spatial}\\
 \chi_i &=& (\frac{1}{2}+\xi_i)\chi_{\uparrow}
 + (\frac{1}{2}-\xi_i)\chi_{\downarrow}.
\end{eqnarray}
$\phi_{{\bf X}_i}$ and $\chi_i$ are spatial and spin functions, respectively, and 
$\tau_i$ is the isospin
function fixed to be up (proton) or down (neutron). 
Accordingly, an AMD wave function
is expressed by a set of variational parameters, ${\bf Z}\equiv 
\{{\bf X}_1,{\bf X}_2,\ldots, {\bf X}_A,\xi_1,\xi_2,\ldots,\xi_A \}$,
which specify centroids of single-nucleon Gaussian wave packets and spin orientations 
for all nucleons.

The parameters ${\bf Z}$ are determined by the energy variation 
after parity and total-angular-momentum projections to obtain the optimized AMD 
wave function.
Namely, in the AMD+VAP method, 
${\bf X}_i$ and $\xi_{i}$($i=1\sim A$) for the lowest $J^\pi$ state 
are determined so as to minimize the energy expectation value of the Hamiltonian,
$\langle \Phi|H|\Phi\rangle/\langle \Phi|\Phi\rangle$,
for the $J^\pi$ eigen wave function projected 
from the AMD wave function; $\Phi= P^{J\pi}_{MK}\Phi_{\rm AMD}({\bf Z})$.
Here, $P^{J\pi}_{MK}$ is the parity and total-angular-momentum projection operator.
For each $J^\pi$ state, the optimum set ${\bf Z}^{(0)}_{J\pi}$ of parameters is obtained. 
After the VAP, to describe 
$J^{\pi}_k$ states, we superpose the $J^{\pi}$-projected AMD wave functions
expressed by the obtained parameter sets ${\bf Z}^{(0)}_{J'\pi'}$ for various $J'^{\pi'}$ as
\begin{eqnarray}
\Psi_{J^{\pi}_k}&=&\sum_{J'^{\pi'}, K} a_{J^{\pi}_k}(J'\pi';K) P^{J'\pi'}_{MK}\Phi_{\rm AMD}({\bf Z}^{(0)}_{J'\pi'}),
\end{eqnarray}
where coefficients $a_{J^{\pi}_k}(J'\pi';K)$ are determined by 
diagonalizing the norm and Hamiltonian matrices. For $Z=N={\rm odd}$ nuclei, 
$T=0$ and $T=1$ projections are approximately done by using the proton-neutron exchanging operator
$P_{p\leftrightarrow n}$ after the energy variation as 
\begin{eqnarray}
\Psi_{J^{\pi}_k}&=&\sum_{J'^{\pi'}, K}\left \{ a_{J^{\pi}_k}(J'\pi';K)+ b_{J^{\pi}_k}(J'\pi';K)
 P_{p\leftrightarrow n}  \right \}\nonumber\\
&& \times  P^{J'\pi'}_{MK}\Phi_{\rm AMD}({\bf Z}^{(0)}_{J'\pi'}),
\end{eqnarray}
where $a_{J^{\pi}_k}(J'\pi';K)$ and  $b_{J^{\pi}_k}(J'\pi';K)$ are determined by 
the diagonalization. For $T=0$ and $T=1$ states, 
$a_{J^{\pi}_k}(J'\pi';K)\approx - b_{J^{\pi}_k}(J'\pi';K)$ and 
$a_{J^{\pi}_k}(J'\pi';K)\approx b_{J^{\pi}_k}(J'\pi';K)$ are obtained, respectively.
In the present framework, we do not explicitly assume
$a_{J^{\pi}_k}(J'\pi';K)=\pm b_{J^{\pi}_k}(J'\pi';K)$ 
because the isospin symmetry is slightly broken in the Hamiltonian
because of  the Coulomb force.
However,  the obtained $\Psi_{J^{\pi}_k}$ for $J^{\pi}_k$ states of $^{10}$B and $^6$Li are found to be approximately $T$ eigen states and can be assigned to
experimental $J^\pi T$ states. 

For the width parameter $\nu$ of single-nucleon Gaussian wave packets,
we choose $\nu=0.235$ fm$^{-2}$ which is the same 
value used for $^{10}$Be and $^{11}$B in Ref.~\cite{Suhara:2009jb}
and was originally determined for $^9$Be in Ref.~\cite{Okabe79}.

In the AMD framework, existence of clusters is not assumed {\it a priori} 
because Gaussian centroids ${\bf X}_i$ of
all single-nucleon wave packets are independently 
treated as variational parameters. 
Nevertheless, if the system energetically favors a specific cluster structure 
such the structure is obtained in the energy variation 
because the AMD model space contains wave functions for various cluster structures. Therefore, the AMD 
method is suitable to investigate 
whether the clusters are formed or not in the system.

Note that the AMD wave function is similar to the wave function used in Fermionic molecular dynamics calculations \cite{Feldmeier:1994he,Neff:2002nu},
though some differences exist in width parameters of single-nucleon Gaussian wave packets and the variational procedure. 
Another difference in the AMD and FMD calculations is effective 
nuclear interaction.  In the AMD calculations, phenomenological effective 
interactions are usually used differently from the recent 
FMD calculations, in which effective interactions
constructed from the realistic nuclear force by means of  
the unitary correlation operator 
method are used \cite{Neff:2002nu}.

\subsection{Effective nuclear interactions}
We use the finite-range central and spin-orbit interactions as effective two-body nuclear interactions,
\begin{eqnarray}
v^{\rm eff}_{12}&=&v_{c}(r)(w+bP_\sigma-hP_\tau-mP_\sigma P_\tau)\nonumber \\
&+& v_{ls}(r) \frac{1+P_\sigma}{2}\frac{1+P_\sigma P_\tau}{2} \bvec{l}\cdot \bvec{s},
\end{eqnarray}
where $P_\sigma$ and $P_\tau$ are the spin and isospin exchange operators, $r$ is the relative distance
$r=|\bvec{r}|$ for the relative coordinate $\bvec{r}=\bvec{r}_1-\bvec{r}_2$,  
$\bvec{l}$ is the angular momentum for $\bvec{r}$, and $\bvec{s}$ is the sum of nucleon spins 
$\bvec{s}=\bvec{s}_1+\bvec{s}_2$. We ignore the $^3$E term of the spin-orbit interaction. 
In the present paper, we use the Volkov No.2 central interaction \cite{VOLKOV},
\begin{eqnarray}
v_{c}(r)&=&v_1\exp\left[-\left(\frac{r}{a_1}\right)^2\right]+v_2\exp\left[-\left(\frac{r}{a_2}\right)^2\right]
\end{eqnarray}
with $v_1=-60.65$ MeV, $v_2=61.14$ MeV,  $a_1=1.80$ fm, and $a_2=1.01$ fm, and
the G3RS spin-orbit interaction \cite{LS},
\begin{eqnarray}
v_{ls}(r)&=&u_1\exp\left[-\left(\frac{r}{b_1}\right)^2\right]+u_2\exp\left[-\left(\frac{r}{b_2}\right)^2\right],
\end{eqnarray}
with $b_1=0.60$ fm and $b_2=0.447$ fm. 

For the Volkov central interaction, we use the Wigner and Majorana parameters, $w=0.40$ and $m=0.60$, which reproduce the $\alpha$-$\alpha$ scattering phase shift, and the Bartlett and Heisenberg parameters, 
$b=h=0.125$ which reproduce the deuteron binding energy. The $b$ and $h$ are the parameters which 
can control the ratio $f$ of the $^3$E interaction to the $^1$E interaction for the fixed 
$w+m$ value as $f=(w+m+b+h)/(w+m-b-h)$. The ratio is $f=1.67$ for the present parametrization. 
Generally, in effective two-body central interactions for structure models, the ratio may change depending on 
nuclear systems because of medium effects and it is usually somewhat suppressed in nuclei. 
Therefore, $b$ and $h$ can be regarded as 
adjustable parameters in nuclei. In addition to the default parametrization $b=h=0.125$, 
we also use a modified one, $b=h=0.06$, which gives a smaller ratio $f=1.27$ to fit the relative 
energy between $T=0$ and $T=1$ states in $^{10}$B spectra. 

For the strengths of the spin-orbit interaction, we take $u_{ls}=u_1=-u_2$.
$u_{ls}$ is the strength parameter of the effective spin-orbit interaction and, in principle, it may depend 
on nuclear systems reflecting contributions from
the three-body force and the tensor force as well as the original spin-orbit force in bare nuclear forces.
It may also have structure model dependence, and therefore, is considered to be an adjustable parameter
in model calculations. In the present paper, we use $u_{ls}=1300$ MeV to reproduce the $ls$ splitting between 
$3/2^-$ and $1/2^-$ states in $^9$Be in the AMD+VAP calculation. We also use a slightly weaker strength
 $u_{ls}=1000$ MeV to see the dependence of energy spectra 
on the strength $u_{ls}$ of the spin-orbit interaction.
The strength of the effective spin-orbit interaction can be estimated by the Scheerbaum factor $B_{S}$
\cite{Kohno:2012vj,Scheerbaum:1976zz}
defined as 
\begin{equation}
B_S=-\frac{2\pi}{q} \int^\infty_0 dr r^3  j_1(qr) v_{ls}(r),
\end{equation}
with $q=0.7$ fm$^{-1}$. Here $j_l$ is the spherical Bessel function. For the G3RS spin-orbit interaction with $u_{ls}=1300$ MeV and 1000 MeV, $B_S$ equals to 103 MeV and 79 MeV.  

In Table \ref{tab:interaction}, we list the adopted interaction parameter sets of effective nuclear interactions
labeled (A) and (B) with the strength $u_{ls}=1300$ MeV and (A') and (B') with $u_{ls}=1000$ MeV. 

\begin{table}[ht]
\caption{
\label{tab:interaction} Adopted parameter sets of effective nuclear interactions.
The Bartlett ($b$) and Heisenberg ($h$) parameters 
for the Volkov No.2 central interaction and the strength parameter $u_{ls}$ for the G3RS spin-orbit interaction.
The Wigner and Majorana parameters are fixed to be $w=0.4$ and $m=0.60$ for all sets.
The ratio of the $^3$E to $^1$E interactions
$f$ of the central interaction and the Scheerbaum factor $B_S$ of the spin-orbit interaction
are also shown.}
\begin{center}
\begin{tabular}{ccccc}
\hline
& (A) & (B) & (A')& (B')\\ 
$b=h$ & 0.125 & 0.06   & 0.125 & 0.06 \\  
$f$ & 1.67 & 1.27 &  1.67 & 1.27\\
$u_{ls}$ [MeV]& 1300 & 1300& 1000& 1000\\
$B_S$ [MeV] & 103 &103 & 79& 79 \\ 
\hline		
\end{tabular}
\end{center}
\end{table}

\section{Results}\label{sec:results}

We calculate $^{10}$B with the AMD+VAP method. 
AMD wave functions for $J^\pi=0^+,1^+,2^+,3^+,4^+$ states are obtained by VAP.
We superpose $J^\pi$-projected states of 10 basis wave functions (five are the obtained wave functions and
five are the $P_{p\leftrightarrow n}$-projected wave functions) to get energy levels. 
We also apply the AMD+VAP method to $^6$Li and $^9$Be and calculate low-lying states,
$^6$Li($1^+$,$2^+$,$3^+$,$0^+$) and $^9$Be($1/2^-,3/2^-,5/2^-,1/2^+,3/2^+,5/2^+$).

\begin{figure}[htb]
\begin{center}
\includegraphics[width=5.5cm]{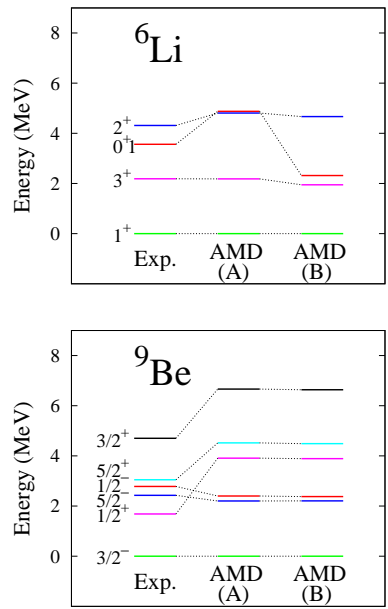} 	
\end{center}
  \caption{(color online). Energy spectra of $^6$Li and $^9$Be obtained by AMD+VAP 
using the interaction parameter sets (A) and
(B) compared with the experimental spectra \cite{AjzenbergSelove:1988ec,Tilley:2002vg,Tilley:2004zz}. 
\label{fig:li6-be9spe}}
\end{figure}

In Fig.~\ref{fig:li6-be9spe}, we show energy spectra of  $^6$Li and $^9$Be obtained by the AMD+VAP calculation
using the interaction parameter sets (A) and (B) compared with the experimental data. 
In the $^6$Li spectra, 
the level spacing between $J^\pi=1^+0$, $3^+0$, and $2^+0$ states is reproduced reasonably.
The excitation energy of the $0^+1$ state is overestimated in the result (A) 
and underestimated in the result (B). This means that
a value of the ratio $f$ in-between $f=1.67$ for (A) and $f=1.27$ for (B) is reasonable to reproduce the $^6$Li spectra. It may indicate
that the effective $^3$E interaction is slightly weaker in $^6$Li than that in a deuteron.
In the $^9$Be spectra, the excitation energy of the $1/2^-$ state is reproduced 
by adjusting the spin-orbit strength $u_{ls}$ as mentioned 
previously. 
Excitation energies of positive-parity states are somewhat overestimated, maybe because
 the present model space of AMD wave functions is not sufficient to describe well 
$K^\pi=1/2^+$ band states, which are
successfully described by molecular orbital models \cite{Okabe79,SEYA}.

\begin{figure}[htb]
\begin{center}
\includegraphics[width=7.8cm]{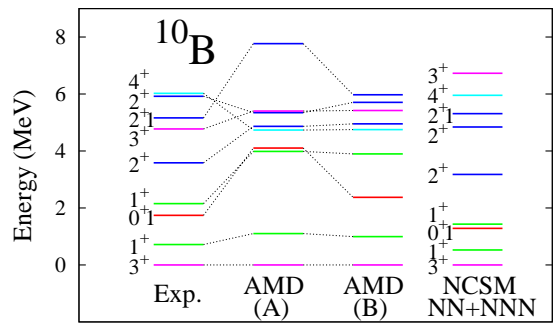} 	
\end{center}
  \caption{(color online). Energy spectra of $^{10}$B. The theoretical result of AMD+VAP
using the interaction parameter sets (A) and (B),
the experimental data \cite{AjzenbergSelove:1988ec,Tilley:2004zz}, and the NCSM calculation 
with the chiral $NN$+$NNN$ force
\cite{Navratil:2007we} are shown.
\label{fig:b10spe}}
\end{figure}

We show the calculated energy spectra of $^{10}$B compared with the 
experimental data in Fig.~\ref{fig:b10spe}.
We also show the energy spectra of the NCSM calculation with the chiral $NN$+$NNN$ force.
Both results (A) and (B) in the present calculation reproduce the ordering of the $3^+_10$ and $1^+_10$ states in 
$^{10}$B. Namely,
the $3^+_10$ is the ground state and the  $1^+_10$ is the first excited state consistently to the
experimental data and also to the NCSM calculation.
The relative energy between the $3^+_10$ and $1^+_10$ states is sensitive to the strength of the effective 
spin-orbit interaction. 
More details of the dependence on the spin-orbit interaction and its relation to the $NNN$ force 
are discussed later. 
The $0^+_11$ energy is largely overestimated in the result (A) and it is reasonably reproduced in the result (B) 
indicating that, in the present model,  the smaller ratio $f\sim 1.27$ of 
the effective $^3$E and $^1$E interactions is favorable for $^{10}$B
than $f\sim 1.67$ for a deuteron.

In Table~\ref{tab:li6-be9-b10}, properties 
of $^6$Li, $^9$Be, and $^{10}$B are listed. The present results are compared with 
the experimental data and also theoretical values of the NCSM calculation with the chiral $NN$+$NNN$ force \cite{Navratil:2007we}. Properties such as radii, moments, and transition strengths are 
reproduced reasonably by the present calculation.

\begin{table}[ht]
\caption{
\label{tab:li6-be9-b10} 
Properties of $^6$Li, $^9$Be, and $^{10}$B. Theoretical values are
calculated by AMD+VAP using the interactions
(A) and (B). 
The experimental proton radii are derived from the charge radii 
in Ref.~\cite{Angeli13}.
Other experimental data are taken from
Refs.~\cite{AjzenbergSelove:1988ec,Tilley:2002vg,Tilley:2004zz}.
The values of the NCSM calculation with the chiral $NN$+$NNN$ force 
from Ref.~\cite{Navratil:2007we} are
also shown.}
\begin{center}
\begin{tabular}{lllll}
\hline
	&	Expt.	&	\multicolumn{2}{c}{AMD+VAP}	&	NCSM	\\
	&		&	(A)	&	(B)	 &	$NN$+$NNN$	\\
\hline
$^6$Li:$|E(1^+_10)|$	&	31.995	&	27.9 	&	26.4 	&	32.63	\\
$r_p(1^+_10)$ [fm]	&	2.44(4)	&	2.21 	&	2.21 	&		\\
$Q(1^+_10)$ [e fm$^2$]	&	$-$0.0818(17)	&	0.09 	&	0.08 	&	$-0.12(4)$	\\
$\mu(1^+_10)$ [$\mu_N$]	&	0.822	&	0.88 	&	0.88 	&	$0.836$	\\
$B(E2;3^+_10\rightarrow 1^+_10)$	&	10.7(8)	&	4.3 	&	4.1 	&	3.685	\\
$B(E2;2^+_10\rightarrow 1^+_10)$	&	4.4(23)	&	5.5 	&	5.2 	&	3.847	\\
$B(M1;0^+_11\rightarrow 1^+_10)$	&	15.4(3)	&	16.1 	&	16.4 	&	15.04(4)	\\
&&&&\\									
$^9$Be:$|E(3/2^-_1)|$	&	58.164	&	53.0 	&	53.0 	&		\\
$r_p(3/2^-_1)$ [fm]	&	2.377(12)	&	2.42 	&	2.42 	&		\\
$Q(3/2^-_1)$ [e fm$^2$]	&	5.288(38)	&	5.2 	&	5.2 	&		\\
$\mu(3/2^-_1)$ [$\mu_N$]	&	$-1.1778(9)$	&	-1.24 	&	-1.24 	&		\\
&&&&\\									
$^{10}$B:$|E(3^+_10)|$	&	64.751	&	58.7 	&	57.7 	&	64.78	\\
$r_p(3^+_10)$ [fm]	&	2.28(5)	&	2.31 	&	2.33 	&	2.197	\\
$Q(3^+_10)$ [e fm$^2$]	&	8.47(6)	&	7.95 	&	8.2 	&	6.327	\\
$\mu(3^+_10)$ [$\mu_N$]	&	1.8006	&	1.84 	&	1.85 	&	1.837	\\
$\mu(1^+_10)$ [$\mu_N$]	&	0.63(12)	&	0.86 	&	0.84 	&		\\
$B(E2;1^+_10\rightarrow 3^+_10)$	&	4.14(2)	&	4.2 	&	3.6 	&	3.05(62)	\\
$B(E2;1^+_20\rightarrow 1^+_10)$	&	15.6(17)	&	10.2 	&	10.1 	&		\\
$B(E2;1^+_20\rightarrow 3^+_10)$	&	1.7(2)	&	0.9 	&	1.3 	&	0.50(50)	\\
$B(E2;2^+_10\rightarrow 1^+_20)$	&	15.2(69)	&	2.7 	&	4.3 	&		\\
$B(E2;2^+_10\rightarrow 1^+_10)$	&	17.8(18)	&	7.9 	&	7.6 	&		\\
$B(E2;2^+_10\rightarrow 3^+_10)$	&	1.2(4)	&	1.1 	&	0.9 	&		\\
$B(E2;3^+_20\rightarrow 1^+_10)$	&	19.7(17)	&	7.7 	&	8.3 	&		\\
$B(M1;0^+_11\rightarrow 1^+_10)$	&	7.5(32)	&	13.5 	&	14.7 	&		\\
$B(M1;1^+_20\rightarrow 0^+_11)$	&	0.19(2)	&	0.0 	&	0.0 	&		\\
$B(M1;2^+_11\rightarrow 2^+_10)$	&	2.5(7)	&	3.7 	&	3.9 	&		\\
$B(M1;2^+_11\rightarrow 1^+_20)$	&	3.1(8)	&	2.8 	&	2.7 	&		\\
$B(M1;2^+_11\rightarrow 1^+_10)$	&	0.32(9)	&	0.2 	&	0.4 	&		\\
\hline		
\end{tabular}
\end{center}
\end{table}

\begin{table}[ht]
\caption{
\label{tab:b10-spin} Expectation values of harmonic 
oscillator quanta and those of the square spin and angular momentum for $^{10}$B
calculated with the interaction (A). For harmonic 
oscillator quanta, the minimum value $Q_{\rm min}=6$ for the $0\hbar\omega$ configuration is subtracted, 
and values of $\Delta Q=\langle Q \rangle -Q_{\rm min}$ are listed.
}
\begin{center}
\begin{tabular}{cccc}
\hline
$^{10}$B($J^\pi T$) &	$\Delta Q$	&	$\langle \bvec{S}^2 \rangle$	&	$\langle \bvec{L}^2 \rangle$	\\
$3^+_10$	&	1.0 	&	2.0 	&	6.8 	\\
$1^+_10$	&	1.5 	&	1.9 	&	0.1 	\\
$0^+_11$	&	0.9 	&	0.5 	&	0.5 	\\
$1^+_20$	&	1.7 	&	1.9 	&	5.7 	\\
$2^+_10$	&	1.4 	&	2.0 	&	6.0 	\\
$3^+_20$	&	1.5 	&	2.0 	&	7.0 	\\
$2^+_11$	&	1.1 	&	0.5 	&	6.0 	\\
$2^+_20$	&	1.5 	&	2.0 	&	6.8 	\\
$4^+_10$	&	1.1 	&	2.0 	&	13.7 	\\
\hline		
\end{tabular}
\end{center}
\end{table}


\section{Discussion}\label{sec:discussion}
\subsection{Dependence of energy spectra on spin-orbit interaction}
\begin{figure}[htb]
\begin{center}
\includegraphics[width=5.5cm]{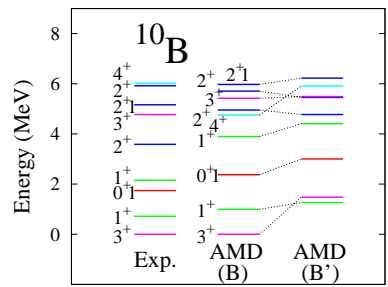} 	
\end{center}
  \caption{(color online). 
Energy spectra for $^{10}$B calculated with AMD+VAP using interactions
(B) and (B'). The experimental data are taken from Refs.~\cite{AjzenbergSelove:1988ec,Tilley:2004zz}
\label{fig:b10-ls-rel}}
\end{figure}

To discuss dependence of the energy spectra 
on the strength of the spin-orbit interaction, we compare the energy spectra obtained using the 
interactions (B) with  the default strength $u_{ls}=1300$ MeV and those obtained using (B') with 
a slightly weak spin-orbit interaction $u_{ls}=1000$ MeV. 
We show $^{10}$B spectra in Fig.~\ref{fig:b10-ls-rel}. 
Energies relative to the $3^+_10$ energy of the result (B) are plotted.
In the result (B') with a weak spin-orbit interaction, the $1^+_10$ energy is lower than 
the $3^+_10$ state as expected from the $pn$ pair picture that an $S=1$ $pn$ pair 
in the $S$-wave is more favored than that in the $D$-wave with no or a weak spin-orbit interaction. 
As a result, the interaction (B') fails to describe the ordering of low-lying energy levels, 
i.e., the ground state spin, $3^+$, of $^{10}$B.
On the other hand, in the result (B), 
the level inversion of the $3^+_10$ and $1^+_10$ states occurs consistently to the experimental data.
The reason for the level inversion is that 
the spin-orbit interaction favors the spin-aligned $T=0$ $pn$ pair 
and lowers the $3^+_10$ state, whereas, it gives almost no contribution 
to the energy of the $T=0$ $pn$ pair  in the $S$-wave in the $1^+_10$ state. 

As for $T=1$ states, the $0^+_11$ state somewhat gains the spin-orbit interaction energy 
because the spin-orbit interaction favors the $T=1$ $pn$ 
pair in the $0^+_1$ state.  In comparison of the results (B) and (B'), it is found that 
the energy gain for the $0^+_1$ state is not as large as 
that for the $3^+_10$ state.

\begin{figure}[htb]
\begin{center}
\includegraphics[width=7.8cm]{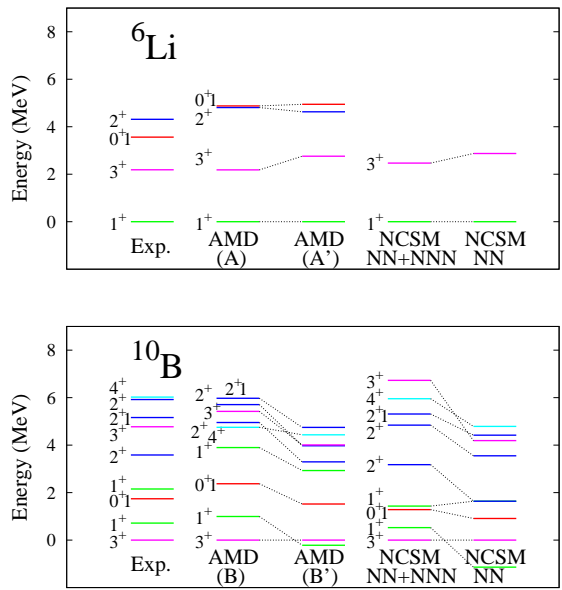} 	
\end{center}
  \caption{(color online). Dependence of the energy spectra 
on the strength $u_{\rm ls}$ of the spin-orbit interaction
for $^{10}$B and $^6$Li calculated with AMD+VAP.
Energy spectra of $^{10}$B obtained using 
(B) with the default spin-orbit interaction 
$u_{\rm ls}=1300$ MeV and (B') with the weaker one
$u_{\rm ls}=1000$ MeV,
and those of $^6$Li obtained using 
(A) with $u_{\rm ls}=1300$ MeV
and (A') with $u_{\rm ls}=1000$ MeV
are shown as well as the experimental energy spectra. 
The NCSM calculation using the chiral nuclear forces 
with the $NNN$ force (the chiral $NN$+$NNN$ force) and without the $NNN$ force (the chiral $NN$ force) from Ref.~\cite{Navratil:2007we} 
are also shown.
\label{fig:b10-li6-ls}}
\end{figure}

In general, a spin-orbit interaction in effective two-body nuclear interactions used for
structure model calculations 
is an effective spin-orbit interaction in nuclei.  
In principle, it should contain a contribution from the $NNN$ force
in addition to the original spin-orbit force in the bare $NN$ forces. 
As Kohno pointed out, the $NNN$ force contributes attractively to the effective 
two-body spin-orbit interaction in nuclear medium \cite{Kohno:2012vj}.
In the G-matrix calculation of nuclear matters 
using the chiral $NN$ and $NN+NNN$ forces in Ref.~\cite{Kohno:2012vj}, 
the contribution of the $NNN$ force is evaluated 
to be $\Delta B_S = 20-30$ MeV of the Scheerbaum factor. 
For instance, in a nuclear matter with the Fermi momentum $k_F=1.35 (1.07)$ fm$^{-1}$, 
the strength is estimated to be $B_S=84.6 (86.5)$ MeV for the chiral $NN$ force
and  $B_S=116.2 (106.7)$ MeV for the chiral $NN+NNN$ force.

In the present calculation, we use the phenomenological effective two-body 
central and spin-orbit interactions, which are adjusted so as to describe the $\alpha$-$\alpha$ scattering 
and the $ls$ splitting in $^9$Be.
Although the present interactions have no direct link to the bare nuclear forces, 
they should indirectly contain the contribution from the $NNN$ force. 
Then, we reach an idea that a part of the two-body spin-orbit interaction in the present effective 
interactions can be interpreted as the contribution from the $NNN$ force. 
With a help of the Kohno's G-matrix analysis, we can roughly 
estimate the contribution of the $NNN$ force 
in the present parametrization as the change $\Delta u_{ls}\sim 300$ MeV of the spin-orbit interaction strength 
which corresponds to the change 
$\Delta B_S = 24$ MeV of the Scheerbaum factor.
Therefore, it is expected that the result (B') with the weaker spin-orbit interaction 
by $\Delta u_{ls}\sim 300$ MeV than the default strength 
can be associated with the calculation without the $NNN$ force contribution
in the effective spin-orbit interaction.
In Fig.~\ref{fig:b10-li6-ls}, we show energy spectra of $^{10}$B 
calculated with the interactions (B) and (B'), and those of the NCSM calculations with the chiral 
$NN+NNN$ and the chiral $NN$ forces. 
In each calculation, the energy of the $3^+_10$ state is set to be zero.
As expected, differences in low-lying spectra between 
results (B) and (B') in the present calculation corresponds well 
to those of the NCSM results with and without the $NNN$ force, 
meaning that the change $\Delta u_{ls}\sim 300$ MeV of the effective two-body spin-orbit interaction
gives effects quite similar to the contribution of the 
$NNN$ force on the low-lying spectra of $^{10}$B. 
For instance, the $1^+_10$ state comes down to the lower energy region 
than the $3^+_10$ state in the result (B') because of the reduction $\Delta u_{ls}\sim 300$ MeV
consistently to the NCSM calculation without the $NNN$ force. 
The excitation energy of the $0^+_11$ state is slightly decreased by the reduction
$\Delta u_{ls}\sim 300$ MeV, which 
corresponds to the difference of the $0^+_11$ excitation energy between 
the NCSM calculation with the $NNN$ force and that without the $NNN$ force.
This association of the present results (B) and (B')
with the NCSM calculations with and without the $NNN$ force
indicates that the part $\Delta u_{ls}\sim 300$ MeV of the two-body spin-orbit interaction 
in the present phenomenological effective interactions is interpreted as 
the contribution of the $NNN$ force, which is essential to the level inversion between 
the $3^+_10$ and $1^+_10$ states in $^{10}$B. 
We also show $^6$Li spectra calculated with 
interactions (A) for the default spin-orbit interaction strength and (A') for the 
reduced strength, compared with the chiral $NN+NNN$ and $NN$ NCSM calculations.
Also for $^6$Li, the change in the low-lying spectra by the reduction of  $\Delta u_{ls}\sim 300$ MeV 
corresponds well to the difference between the NCSM calculations with and without the 
$NNN$ force. 

\subsection{Structure of $^{10}$B}
We analyze $^{10}$B wave functions obtained by AMD+VAP
and find that the ground and excited states of $^{10}$B are approximately understood
by $T=0$ of $T=1$ $pn$ pairs around the 2$\alpha$ core.
In Table \ref{tab:b10-spin}, we show expectation values of the squared intrinsic spin,  
$\langle \bvec{S}^2 \rangle$, and those of the squared orbital angular
momentum, $\langle \bvec{L}^2 \rangle$. We also show expectation values 
of the harmonic oscillator quanta,  $\langle Q \rangle$, 
given by the creation and annihilation operators $Q=a^\dagger a$ 
of the harmonic oscillator for the width parameter $\nu=0.235$ fm$^{-2}$. 
Since the $2\alpha$ core gives no contribution to the total intrinsic spin, 
$\langle \bvec{S}^2 \rangle$ reflects mainly intrinsic spin configurations of 
two nucleons around the core. The calculated values of
 $\langle \bvec{S}^2 \rangle$ for $T=0$ states are $\langle \bvec{S}^2 \rangle\approx 2$
indicating that two nucleons form a $(ST)=(10)$ pair, which is the same spin-isospin 
configuration as a deuteron.  For $T=1$ states,
$\langle \bvec{S}^2 \rangle$ is approximately 0.5 meaning that the $T=1$ $pn$ pair  has
the dominant $(ST)=(01)$ component with a mixing of $S=1$ component.
The $S=1$ mixing in the $T=1$ $pn$ pair is nothing but the odd-parity mixing
in the pair caused by the spin-orbit potential from the core as discussed in the 
previous paper for the $pn$ pair around the $^{16}$O core in $^{18}$F.
$\langle Q \rangle$ for the $1^+_10$ state is relatively large compared with 
those for the $3^+_10$ and $0^+_11$ states because the  $1^+_10$ state has 
a spatially developed $pn$ pair as well as the $2\alpha$ clustering and contains 
higher shell components.

\begin{figure}[htb]
\begin{center}
\includegraphics[width=7.8cm]{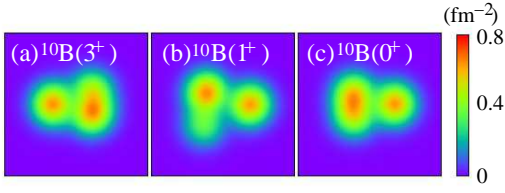} 	
\end{center}
  \caption{(color online). Distributions of matter densities of $^{10}$B$(3^+_1)$, 
$^{10}$B$(1^+_1)$, and $^{10}$B$(0^+_1,T=1)$
calculated with AMD+VAP using the interaction (A). 
Densities of intrinsic states are integrated with respect to the $z$ axis and
plotted on the $x$-$y$ plane (the box size is 10 fm $\times$ 10 fm). 
Here, axes of the intrinsic frame are chosen as 
$\langle x^2\rangle \ge \langle y^2\rangle \ge \langle z^2\rangle$. 
\label{fig:b10-dense}}
\end{figure}

Figure \ref{fig:b10-dense} shows the matter density 
distribution of the intrinsic wave functions 
 for the $3^+_10$, $1^+_10$, and $0^+_11$ states. 
The density of the single AMD wave function obtained by VAP for each
$J^\pi$ is shown. 
In the $3^+_10$ state, the $T=0$ $pn$ pair exists at the surface of an $\alpha$ cluster, whereas,
in the $1^+_10$ state, it spatially develops.
In the $0^+_11$ state, the $T=1$ $pn$ pair locates close to an $\alpha$ cluster.  
As mentioned previously, the $T=0$ $pn$ pair in the $3^+_10$ state 
and the $T=1$ $pn$ pair in the $0^+_11$
state are energetically favored by the spin-orbit potential from the core.
To gain the spin-orbit potential, the $pn$ pair
remains at the surface close to the core in the $3^+_10$ and $0^+_11$ states.
This is contrast to the spatially developed $T=0$ $pn$ pair in the $1^+_10$ state,
in which the spin-orbit interaction gives minor contribution. 

\section{$2\alpha+pn$ model  analysis of $pn$ pair} \label{sec:2alpha-pn}

As discussed previously, 
the $3^+_10$  and  $0^+_11$ states gain the spin-orbit interaction, 
whereas the $1^+_10$ state is not affected by the spin-orbit interaction.
This result is understood by effects of the spin-orbit potential to $T=0$ and $T=1$ $pn$ pairs 
at the nuclear surface, which were
discussed in the previous paper for $^{18}$F based on
the $^{16}$O+$pn$ model. 
To reveal the role of the spin-orbit interaction in the $^{10}$B system, 
we here apply a $2\alpha+pn$ model 
and investigate effects of the spin-orbit interaction to the $pn$ pair
at the surface of the $2\alpha$ core. 

\begin{figure}[htb]
\begin{center}
\includegraphics[width=7cm]{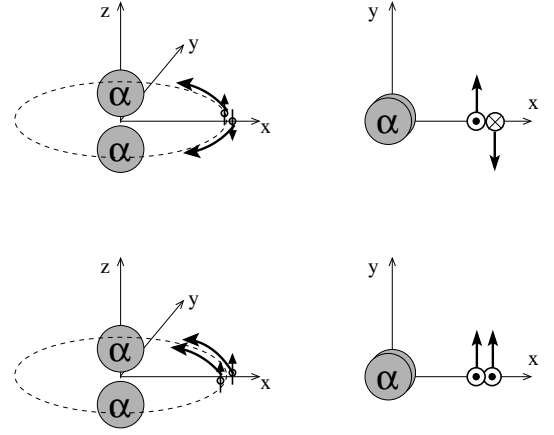} 	
\end{center}
  \caption{Schematic figures for the $S_z=0$ $pn$ pair (upper figures) and 
the $S_z=1$ $pn$ pair (lower figures) around 
the $2\alpha$ core in the $2\alpha+pn$ model.
\label{fig:2alpha-pn}}
\end{figure}

\begin{figure}[htb]
\begin{center}
\includegraphics[width=5cm]{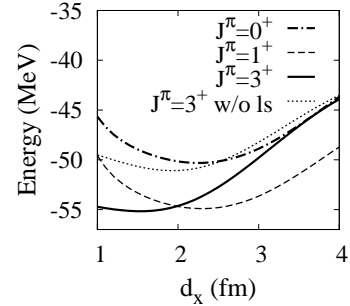} 	
\end{center}
  \caption{$d_x$ dependence of energy calculated with the $2\alpha+pn$ model.
Energies of $J^\pi=0^+$, $1^+$ and $3^+$ projected states for $k_y=0$
 are calculated using the interaction (A). 
The $\alpha$-$\alpha$ distance is fixed to be $d_{\alpha\alpha}=3$ fm.
$S_z=0$ is chosen for the $0^+$ state, and  
$S_z=1$ and $K=1$($K=3$) are chosen for the $1^+$($3^+$) state.
The $3^+$ energy calculated 
without the spin-orbit interaction is also shown.
\label{fig:be8-2N-d}}
\end{figure}

\begin{figure}[htb]
\begin{center}
\includegraphics[width=5cm]{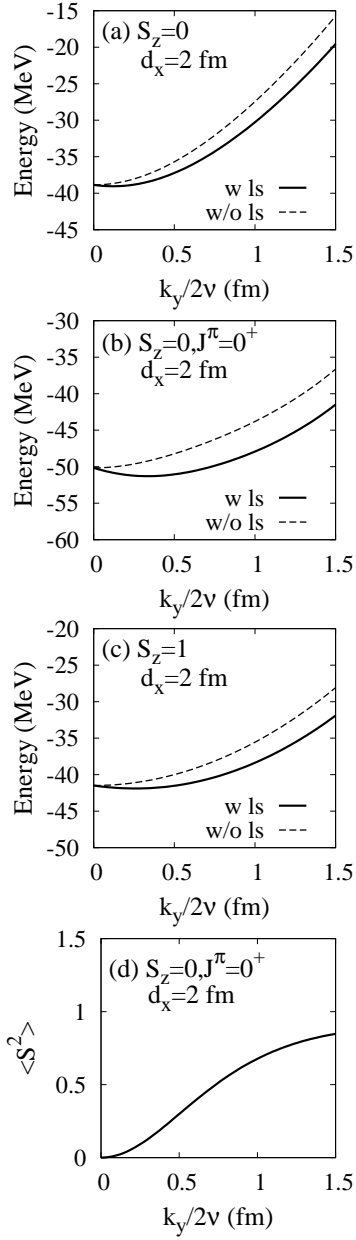} 	
\end{center}
\caption{Energies and $\langle S^2 \rangle$ calculated with the $2\alpha+pn$ model
using the interaction (A).
(a) Energy of the non-projected state
for the $S_z=0$ $pn$ pair, (b) that of the $J^\pi=0^+$ projected state
for the $S_z=0$ $pn$ pair, and (c) that of the non-projected state for
the $S_z=1$ $pn$ pair. The energies with and without
the spin-orbit force are plotted as functions of $k_y$.
$d_{\alpha\alpha}=3$ fm and $d_x=2$ fm are chosen. 
(d) The spin expectation value $\langle S^2 \rangle$ of the $J^\pi=0^+$ projected state
for the $S_z=0$ $pn$ pair.
\label{fig:be8-2N-k}}
\end{figure}

\begin{figure}[htb]
\begin{center}
\includegraphics[width=5cm]{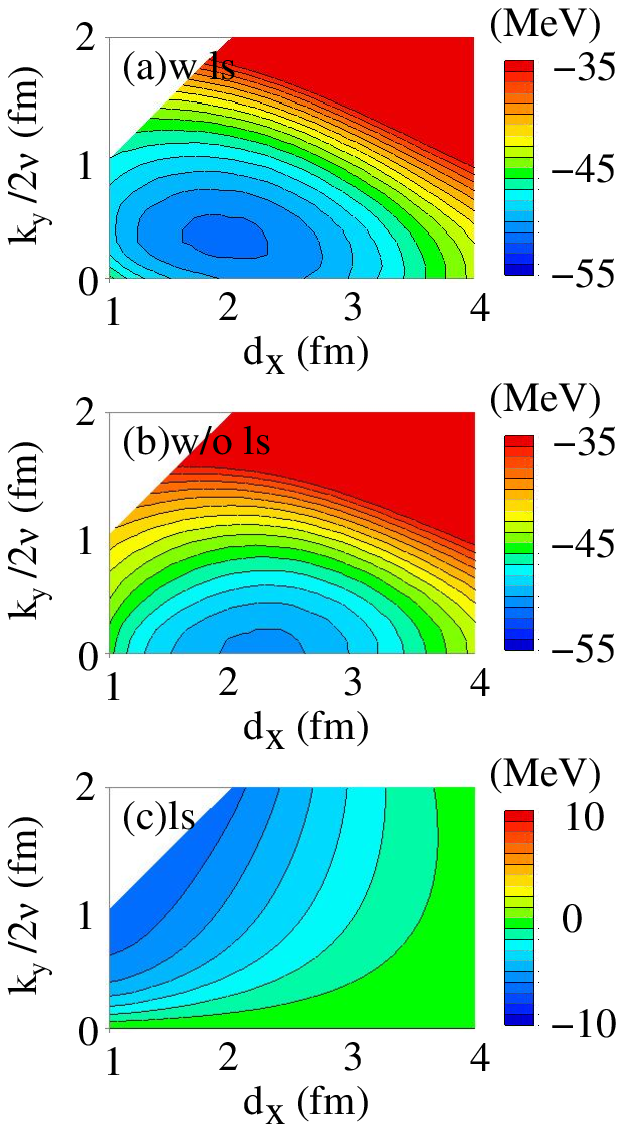} 	
\end{center}
  \caption{(color online). Energies of the $J^\pi=0^+$ projected state 
for the $S_z=0$ $pn$ pair calculated with the $2\alpha+pn$ model using 
the interaction (A).
(a) Energy with the spin-orbit interaction, 
(b) that without the spin-orbit interaction, and (c) the expectation value of the 
spin-orbit interaction.
\label{fig:d-di}}
\end{figure}

Let us consider a proton and a neutron at the surface of the $2\alpha$ core. 
Because of the $^3$E and $^1$E interactions, they form $(ST)=(10)$ 
and $(ST)=(01)$ pairs. The former is the deuteron-like $pn$ pair and 
the latter corresponds to the dineutron pair. For simplicity, 
we 
consider two nucleons with parallel intrinsic spins for the
$T=0$ pair and untiparallel intrinsic spins for the $T=1$ pair 
around the $2\alpha$ core 
as shown
in Fig.~\ref{fig:2alpha-pn}.
Here we take intrinsic spin orientations along the $z$-axis for the $\alpha$-$\alpha$ direction.
Without the spin-orbit potential from the core, it is naively expected that  
$T=0$ and $T=1$ pairs move 
in the $S$-wave ($L=0$) around the $2\alpha$ core
in the lowest state to construct 
$J^\pi T=1^+0$ and $0^+1$ states. Due to the stronger
$^3$E interaction than the $^1$E interaction, 
the $1^+0$ state is expected to be lower than the $0^+1$ state.
In the spin-orbit potential from the core, a spin-up nucleon at the surface 
is boosted to have finite momentum and a spin-down nucleon is boosted to the 
opposite direction. Consequently, for the $(ST)=(10)$ pair, the spin-orbit potential
boosts two nucleons in the 
same direction and causes the orbital rotation of the pair, and therefore it favors 
the spin-aligned $J^\pi T=3^+0$ state.
For the $T=1$ pair, the spin-orbit potential boosts
two nucleons in the opposite direction. Due to the opposite boosting by the spin-orbit potential,
the $T=1$ pair is no longer the ideal $(ST)=(01)$ pair  but it contains 
the odd-parity mixing, i.e., the mixing of the $S=1$ component in the dominant $S=0$ component
as discussed in the previous paper.

To quantitatively discuss contributions of the spin-orbit interaction to 
$T=0$ and $T=1$ pairs in the $2\alpha+pn$ system, 
we introduce a $2\alpha+pn$ model as follows. 
The $2\alpha+pn$ wave function with anti-parallel spins ($S_z=0$)  for the $T=1$ $pn$ pair  is given as 
\begin{eqnarray}
&&\Phi^{S_z=0}_{2\alpha+pn}= {\cal A} 
\left\{ 
\Phi_{\alpha}(\bvec{R}_{1})\Phi_{\alpha}(\bvec{R}_{2})
\psi_{p\uparrow}(\bvec{X}_1) \psi_{n\downarrow}(\bvec{X}_2) \right\},\label{eq:s0-pn}\\
&& \psi_{\tau\sigma}(\bvec{X};\bvec{r}) 
=  \phi_{\bvec{X}}(\bvec{r}) \chi_{\tau\sigma},
\end{eqnarray}
where $\Phi_{\alpha}(\bvec{R}_{k})$ is the $\alpha$ cluster wave
function written by the $(0s)^4$ harmonic oscillator configuration located at $\bvec{R}_{k}$, 
and $\psi_{\tau\sigma}$ is the single-particle wave function for a valence nucleon assumed to be a localized Gaussian
wave packet. Here we use 
labels $\tau=p,n$ and $\sigma=\uparrow,\downarrow$ for the isospin and intrinsic spin of the nucleon,  
respectively.
We set two $\alpha$s with the distance $d_{\alpha\alpha}$ parallel to the $z$-axis
as $\bvec{R}_{1}=-\bvec{R}_{2}=(0,0,d_{\alpha\alpha}/2)$,
and the single-nucleon Gaussian wave packets for $p\uparrow$ and $n\downarrow$ at 
\begin{eqnarray}
&&\bvec{X}_1=(d_x,i k_y/2\nu,0),\\
&&\bvec{X}_2=(d_x,-i k_y/2\nu,0).
\end{eqnarray}
Here, parameters $d_x$ and $k_y$ stand for the mean positions and momenta
of the Gaussian wave packets,
\begin{eqnarray}
&&\langle \phi(\bvec{X}_{1,2})| \hat{\bvec{r}} |\phi(\bvec{X}_{1,2})\rangle=  (d_x,0,0),\\
&&\langle \phi(\bvec{X}_1)| \hat{\bvec{p}} |\phi(\bvec{X}_1)\rangle= (0,\hbar k_y,0),\\
&&\langle \phi(\bvec{X}_2)| \hat{\bvec{p}} |\phi(\bvec{X}_2)\rangle= (0,-\hbar k_y,0),
\end{eqnarray}
meaning that spin-up and -down nucleons are boosted to have 
finite momenta in the opposite direction
(see upper panels of Fig.~\ref{fig:2alpha-pn}).
This parametrization is a kind of extension of the model for $\alpha$ cluster structures 
proposed by Itagaki {\it et al.} in Ref.~\cite{Itagaki:2005sy}. 
Note that,  in the $k_y\ne 0$ case,  the $pn$ pair contains the $S=1$ component in addition to
the dominant $S=0$ component.
The $2\alpha+pn$ wave function with parallel spins ($S_z=1$) for the $T=0$ $pn$ pair is written as 
\begin{eqnarray}
&&\Phi^{S_z=1}_{2\alpha+pn}= {\cal A} 
\left\{ 
\Phi_{\alpha}(\bvec{R}_{1})\Phi_{\alpha}(\bvec{R}_{2})
\psi_{p\uparrow}(\bvec{X}_1) \psi_{n\uparrow}(\bvec{X}_2) \right\},\label{eq:s1-pn},
\end{eqnarray}
with 
\begin{eqnarray}
&&\bvec{X}_1=(d_x,i k_y/2\nu,0),\\
&&\bvec{X}_2=(d_x,i k_y/2\nu,0),
\end{eqnarray}
where nucleons in the $pn$ pair are boosted in the same direction 
(see lower panels of Fig.~\ref{fig:2alpha-pn}). 

For simplicity we fix the $\alpha$-$\alpha$ distance as $d_{\alpha\alpha}=3$ fm. 
The contribution from the center of mass motion is exactly removed 
by shifting Gaussian center positions as 
$\bvec{R}_{1, 2}\rightarrow \bvec{R}_{1,  2}-\bvec{R}_G$ and 
$\bvec{X}_{1,2}\rightarrow \bvec{X}_{1,2}-\bvec{R}_G$ with
\begin{equation}
\bvec{R}_G=\frac{4(\bvec{R}_{1}+\bvec{R}_{2})+\bvec{X}_1+\bvec{X}_2}{10}.
\end{equation}
The $J^\pi$ state projected from $\Phi^{S_z=\{0,1\}}_{2\alpha+pn}$ is given as
\begin{equation}
|J^\pi M\rangle_K=P^{J\pm}_{MK}\Phi^{S_z=\{0,1\}}_{2\alpha+pn}.
\end{equation}
We calculate energy expectation values of the $2\alpha$+$pn$ wave functions using the 
interaction (A) and that without the spin-orbit interaction, and analyze energies of 
the $T=1$ and $T=0$ $pn$ pairs in the $2\alpha+pn$ system. 

We first discuss energies of $2\alpha+pn$ for the $k_y=0$ case with no boosting
which corresponds to ideal $(ST)=(01)$ and $(ST)=(10)$ $pn$ pairs. 
Figure \ref{fig:be8-2N-d} shows energies of 
the $J^\pi=0^+$, $1^+$ and $3^+$ projected states, 
$P^{0+}_{M0}\Phi^{S_z=0}_{2\alpha+pn}$, 
$P^{1+}_{M1}\Phi^{S_z=1}_{2\alpha+pn}$, and  
$P^{3+}_{M3}\Phi^{S_z=1}_{2\alpha+pn}$, 
plotted 
as functions of the distance $d_x$ of the pair position from the core.
Here, $K=0$, $K=1$, and $K=3$ are chosen for 
$J^\pi=0^+$, $1^+$ and $3^+$ projections, respectively.
Note that the $J^\pi=0^+$ projected wave function is a $T=1$ eigen state and
$\Phi^{S_z=1}_{2\alpha+pn}$ is a $T=0$ eigen state.
The $J^\pi=0^+$ and $1^+$ energy curves have minimums
in the $d_x > 2$ fm region indicating that the ideal $T=1$ and $T=0$ $pn$ pairs
develop spatially from the core. The spin-orbit interaction gives no contribution to 
the $(ST)=(01)$ pair in the $0^+$ state nor to the $(ST)=(10)$
pair in the $1^+$ state.
In the $3^+$ energy curve obtained without the spin-orbit interaction, 
the optimum $d_x$ at the energy minimum is slightly smaller than 
those for the $1^+$ and $0^+$ energy curves because of the relatively high centrifugal 
barrier. The $3^+$ energy obtained with the spin-orbit interaction shows 
a large energy gain in the small $d_x$ region. It indicates that 
the spin-aligned $T=0$ pair is favored by the spin-orbit potential from the core, 
which keeps the pair close to the core. 

Next we analyze the $k_y\ne 0$ case to discuss the contribution of 
the nucleon momenta. 
Figure \ref{fig:be8-2N-k} shows intrinsic
energies of the ${2\alpha+pn}$
wave functions for $S_z=0$ and $S_z=1$  without the $J^\pi$ projection
and the $J^\pi=0^+$ projected energy
for $S_z=0$. Energies are plotted as functions of the momentum $k_y$.
For the $J^\pi=0^+$ projected state, we also show 
$\langle\bvec{S}^2\rangle$, which indicates the $S=1$ mixing (the odd-parity mixing)
in the $S=0$ component as a function of $k_y$.
The $pn$ pair  position $d_x$ is fixed to be $d_x=2$ fm.
In Figs.~\ref{fig:be8-2N-k}(a) and (c) for intrinsic energies, it is found that intrinsic states gain the 
spin-orbit interaction in the finite $k_y$ region because of the 
boosting of nucleons in the opposite directions in the $S_z=0$ pair and 
that in the same direction in the $S_z=1$ pair. 
In the energy curve for the $J^\pi=0^+$-projected state (see Fig.~\ref{fig:be8-2N-k} (b)),
a further large energy gain of the spin-orbit interaction is found in the finite $k_y$ region.

In Fig.~\ref{fig:d-di}, we show the $0^+$ energy 
with and without the spin-orbit interaction plotted on the $d_x$-$k_y$ plane. We also 
show the expectation value of the spin-orbit interaction of the $0^+$ projected state.
The energy surface obtained without the spin-orbit interaction shows the 
energy minimum at $d_x= 2.2$ fm on the $k_y=0$ line (see Fig.~\ref{fig:d-di}(b)). 
The contribution of the spin-orbit interaction is attractive 
in the finite $k_y$, in particular, in the small $d_x$ region  (see Fig.~\ref{fig:d-di}(c)), in which
two nucleons in the $S_z=0$ pair approximately occupy 
the single-particle $|\Omega|=|j_z|=3/2$ orbits in the $p$ shell.
Consequently, the energy minimum shifts to the finite $k_y$ 
and slightly smaller $d_x$ region in the result with the spin-orbit interaction  (see Fig.~\ref{fig:d-di}(a)).
This result indicates that the $(ST)=(01)$ $pn$ pair in the $0^+$ state is somewhat broken to 
contain the odd-parity mixing (the $S=1$ mixing in the $S=0$ component) 
by the spin-orbit potential at the surface from the core. Moreover, 
because of the spin-orbit potential, the spatial development of the 
$pn$ pair is suppressed slightly.

\section{Summary} \label{sec:summary}
We investigated the structures of positive-parity states of $^{10}$B 
with AMD+VAP using the phenomenological effective 
two-body interactions. In the result, we found 
$2\alpha+pn$ structures in $^{10}$B.
We discuss effects of the spin-orbit force 
on the energy spectra and $pn$ correlations in the $J^\pi T=1^+_10$, $3^+_10$, and 
$0^+_11$ states.
The $1^+_10$ state is not affected by the spin-orbit interaction,
whereas the $3^+_10$ state gains energy of the spin-orbit interaction largely
to come down to the ground state, and  the $0^+_11$ state also gains somewhat energy of the 
spin-orbit interaction.
We showed that the change $\Delta u_{ls}\sim 300$ MeV of the spin-orbit interaction in the present 
effective two-body interactions gives effects quite similar to the contribution of the 
$NNN$ force in the NCSM calculation on the low-lying spectra of $^{10}$B and $^6$Li. 
It indicates that the part of the two-body spin-orbit interaction 
can be interpreted as a contribution of the $NNN$ force, which is essential to the level ordering of 
the $3^+_10$ and $1^+_10$ states in $^{10}$B. 
We also applied the $2\alpha+pn$ model and discuss the effects of the spin-orbit interaction on
the $T=0$ and $T=1$ $pn$ pairs around the 2$\alpha$ core.
In the spin-aligned $J^\pi T=3^+0$ state, the spin-orbit interaction affects the $(ST)=(10)$ pair attractively
and suppresses the spatial development of the pair,  
whereas, in the $1^+0$ state, it gives a minor effect to the $(ST)=(10)$ pair.
The $(ST)=(01)$ pair in the $0^+1$ state is somewhat dissociated to have the odd-parity mixing, 
i.e., the mixing of $S=1$ component by the spin-orbit 
interaction. 

In the present calculation, we use the phenomenological effective two-body 
central and spin-orbit interactions, which are adjusted so as to describe the $\alpha$-$\alpha$ scattering 
and the $ls$ splitting in $^9$Be.
The present interactions have no direct link to the bare nuclear force 
although  the contributions from the $NNN$ force as well as the tensor force 
and also many-body effects in nuclear systems should be 
indirectly contained in the effective interactions.
In the present paper, we associate a contribution of the $NNN$ force with a part of the 
effective two-body spin-orbit interaction with the help of the G-matrix calculation by Kohno. 
It is a remaining future problem to adopt more sophisticated effective interactions 
derived from bare nuclear forces and investigate effects of the $NNN$ force on the 
$pn$ correlations in $Z=N={\rm odd}$ nuclei. 
It is also an important issue to study effects of the $NNN$ force 
on nuclear structures considering the link of the $NNN$ force with the effective two-body
spin-orbit interactions as done for nuclear radii by Nakada {\it et al.} \cite{Nakada:2014apa}.

\section*{Acknowledgments} 
The computational calculations of this work were performed by using the
supercomputers at YITP. This work was supported by 
JSPS KAKENHI Grant Numbers 26400270 and 14J02221.

\end{document}